\documentclass{ws-rv9x6}
\usepackage{ws-rv-vanmod}             % numbered citation/references (default)
\makeindex
\newcommand\TL{\hfil$\displaystyle{##}$}
\newcommand\TR{$\displaystyle{{}##}$\hfil}
\newcommand\TC{\hfil$\displaystyle{##}$\hfil}
\newcommand\TT{\hbox{##}}

\def\seqalign#1#2{\vcenter{\openup1\jot
  \halign{\strut #1\cr #2 \cr}}}

\def\mop#1{\mathop{\rm #1}\nolimits}

\def\diag{\mop{diag}}
\def\tr{\mop{tr}}

\def\arcsinh{\mop{arcsinh}}
\def\arctanh{\mop{arctanh}}
\def\slashed#1{{\ooalign{\hfil\hfil/\hfil\cr $#1$}}}
\def\lbldef#1#2{\expandafter\gdef\csname #1\endcsname {#2}}
\newcommand{\eqn}[3][]{\lbldef{#2}{(\ref{#2})}%
\def\@eqnstyle{#1}%
\ifx\@eqnstyle\@empty%
\begin{equation} #3 \label{#2} \end{equation}%
\else%
\begin{equation} \seqalign{\span\TC}{#3} \label{#2} \end{equation}%
\fi}
\def\eqalign#1{\vcenter{\openup1\jot
    \halign{\strut\span\TL & \span\TR\cr #1 \cr
   }}}
\def\eno#1{(\ref{#1})}
\begin{document}

\chapter{TASI lectures: Collisions in anti-de Sitter space, conformal symmetry, and holographic superconductors}

\author[S. Gubser]{Steven S. Gubser}

\address{Joseph Henry Laboratories \\
Princeton, NJ  08540 \\
ssgubser@princeton.edu}

\begin{picture}(0,0)(0,0)
\put(258,200){PUPT-2362}
\end{picture}
\vskip-0.17in
\begin{abstract}
In four lectures, delivered at the TASI 2010 summer school, I cover selected topics in the application of the gauge-string duality to nuclear and condensed matter physics.  On the nuclear side, I focus on multiplicity estimates from trapped surfaces in $AdS_5$, and on the consequences of conformal symmetry for relativistic hydrodynamics.  On the condensed matter side, I explain the fermion response to the zero-temperature limit of $p$-wave holographic superconductors.
\end{abstract}

\body

\section{Introduction}
\label{INTRODUCTION}

In my TASI 2010 lectures, I discussed two developments in applications of the gauge-string duality.\cite{Maldacena:1997re,Gubser:1998bc,Witten:1998qj}  The first is aimed at nuclear physics and focuses on multiplicity estimates from trapped surfaces and $O(3)$ symmetry from colliding black holes in $AdS_5$.  The second is aimed at condensed matter physics and treats $p$-wave superconductors and the response of fermions to them from holographic duals.

My lectures were among the last at the school, and they were intended as ``special topics'' lectures.  As a result, I took more time than usual to explain how my own understanding of these subjects developed, why I worked on them, and what questions I was asking myself at the time.  I did this in the hope that students would ponder whether my approach to sniffing out research problems had some relevance for them.

This writeup adheres closely to the order of presentation in my actual lectures.  In section~\ref{NUCLEAR} (Lectures 1 and 2) I discuss black hole collisions in $AdS_5$, trapped surfaces, and $O(3)$ symmetry.  These lectures are based on work done in part with S.~Pufu and A.~Yarom.\cite{Gubser:2008pc,Gubser:2010ze,Gubser:2010ui}  In sections~\ref{PWAVE} and~\ref{FERMIONS} (Lectures 3 and 4), I turn to a discussion of the fermion response to $p$-wave holographic superconductors, based on work done with F.~Rocha and A.~Yarom.\cite{Gubser:2010dm}

\section{Lecture 1: Trapped surfaces in $AdS_5$ and $O(3)$ symmetry}
\label{NUCLEAR}

\subsection{Overview of the main results}

When pointlike, lightlike particles collide head-on in $AdS_5$, a black hole forms with
 \eqn{SBH}{
  S_{\rm BH} \geq S_{\rm trapped} \approx \pi \left( {L^3 \over G_5} \right)^{1/3}
    (2EL)^{2/3} \,,
 }
where $E$ is the energy of {\it one} of the particles, and we assume $EL \gg 1$.\cite{Gubser:2008pc}.  The main meat of my first lecture was to explain how \eno{SBH} comes about, and to draw attention to an $O(3)$ symmetry that comes up along the way.  But first I will provide an overview of what the various quantities in \eno{SBH} mean and why it seemed to me a good idea to work out the inequality \eno{SBH}.

\begin{figure}
\centerline{\psfig{file=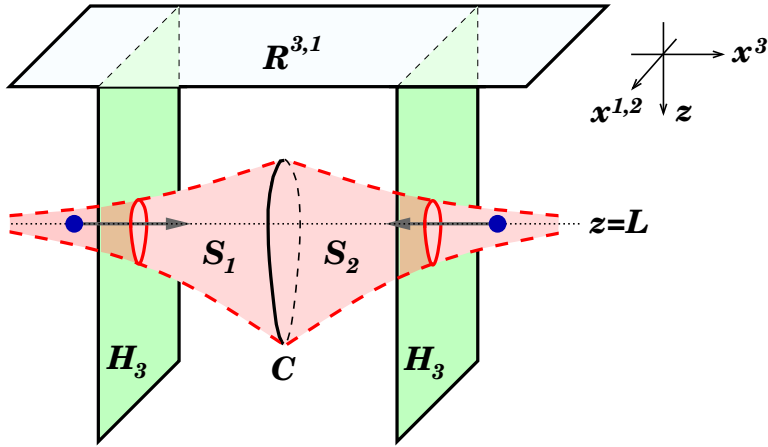,width=4in}}
\caption{A rough depiction of the trapped surface in $AdS_5$ formed in a collision of massless particles.  From the original literature.\cite{Gubser:2008pc}}\label{3Dtrapped}
\end{figure}

$S_{\rm BH} = A/4G_5$ is the Bekenstein-Hawking entropy.  $G_5$ is Newton's constant in $AdS_5$.  $L$ is the radius of curvature of $AdS_5$, whose Ricci tensor takes the form
 \eqn{Rmn}{
  R_{\mu\nu} = -{4 \over L^2} g_{\mu\nu} \,.
 }
$S_{\rm trapped}$ is the area of a trapped surface: a closed, spacelike, co-dimension two surface in $AdS_5$ both of whose forward-directed normal vectors point inward.  A rough depiction of the trapped surface is shown in figure~\ref{3Dtrapped}.

Plugging numbers into \eno{SBH} which are suitable for comparison to a top-energy gold-gold collision at RHIC ($\sqrt{s_{NN}} = 200\,{\rm GeV}$) gives $S_{\rm trapped} \approx 35,\!000$, quite close to phenomenological estimates $S_{\rm Au\,Au} \approx 38,\!000$ for central collisions.  But data at RHIC energies and below favors a slower scaling with beam energy $E$, namely $S \propto E^\alpha$ where $\alpha \approx 1/2$ or a bit smaller: see for example the discussion by Steinberg.\cite{Steinberg:2004vy}

Now, why work on trapped surfaces in $AdS_5$?  In no particular order, here are the reasons that I recall as motivations.
 \begin{itemize}
  \item I had read work by Eardley and Giddings\cite{Eardley:2002re} treating the possibility of black hole formation in $pp$ collisions at the LHC.
  \item The idea of quantitatively comparing black holes in $AdS_5$ to the quark-gluon plasma (QGP) was well-established.
  \item Formation of the QGP was (and is) recognized as a hard and interesting problem.
  \item Black hole formation is interesting on formal grounds, and trapped surfaces provide a standard first cut at the problem.
  \item Penrose argued in unpublished work that trapped surfaces have to be entirely enclosed by a black hole horizon.
 \end{itemize}

In explaining a derivation of \eno{SBH}, I will inevitably leave out a fair amount of detail.  Readers interested in seeing the details are referred to the original literature.\cite{Gubser:2008pc}

\subsection{Shock waves in $AdS_5$}
\label{SHOCKS}

Before the collision of lightlike particles, the geometry we want to consider (including the back-reaction from the massless particles) is
 \eqn{dsTwo}{
  ds^2 = {L^2 \over z^2} \left[ -du dv + (dx^1)^2 + (dx^2)^2 + dz^2 \right] + 
    \delta ds^2
 }
where
 \eqn{ddsTwo}{
  \delta ds^2 = {L \over z} \Phi(x^1,x^2,z) \left[ \delta(u) du^2 + \delta(v) dv^2 \right]
 }
and
 \eqn{uAndv}{
  u = t - x^3 \qquad v = t + x^3 \,.
 }
The scalar function $\Phi$ takes the form
 \eqn{PhiForm}{
  \Phi = {2G_5 E \over L} {1 + 8q(1+q) - 4 \sqrt{q(1+q)} (1+2q) \over \sqrt{q(1+q)}}
 }
where
 \eqn{qForm}{
  q = {(x^1)^2 + (x^2)^2 + (z-L)^2 \over 4zL} \,.
 }
As we will see in section~\ref{OTHREE}, $q$ is essentially the only combination of $x^1$, $x^2$, and $z$ that respects an $O(3)$ symmetry which preserves the worldlines of the massless particles prior to the collision.

The stress energy tensor dual to the shock wave metric is
 \eqn{TuuAndTvv}{\eqalign{
  \langle T_{uu} \rangle &= {2L^4 E \over \pi \left[ L^2 + (x^1)^2 + (x^2)^2 \right]^3}
     \delta(u)  \cr
  \langle T_{vv} \rangle &= {2L^4 E \over \pi \left[ L^2 + (x^1)^2 + (x^2)^2 \right]^3}
     \delta(v) \,,
 }}
with all other components vanishing.

The shock wave metric \eno{dsTwo} is an {\it exact} solution to
 \eqn{SourcedEinstein}{
  R_{\mu\nu} = -{4 \over L^2} g_{\mu\nu} + \hbox{(massless pointlike sources)}
 }
outside the causal future of the collision plane $u=v=0$.  {\it Inside} this causal future, i.e.~for $u$ and $v$ positive, it's hard to compute the metric: all the difficulties of classical black hole formation live here.

\begin{figure}
\centerline{\psfig{file=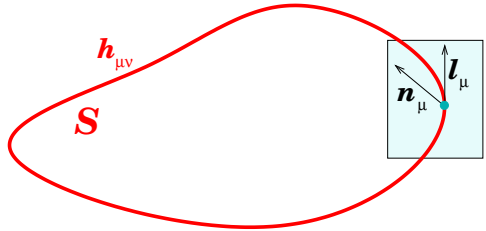,width=2.5in}}
\caption{A trapped surface ${\bf S}$ with induced metric $h_{\mu\nu}$ and null normal vectors $(n^\mu,\ell^\mu)$.  Although the vector field $\ell^\mu$ points more outward than $n^\mu$, its expansion is negative.}\label{TrappedSurface}
\end{figure}

Although it would be possible to spend considerably more space explaining where the results \eno{PhiForm} and \eno{TuuAndTvv} come from, let me pass on instead to trapped surfaces.  Because a trapped surface ${\bf S}$ is spacelike and co-dimension $2$, there is a ``normal plane'' at each point along it, spanned by one timelike and one spacelike vector.  More conveniently, as shown in figure~\ref{TrappedSurface}, let $(n^\mu,\ell^\mu)$ be a null basis for the normal plane, with both $n^\mu$ and $\ell^\mu$ future-directed.  Let $n^\mu$ be the more inward-pointing of the two null basis vectors.  If $h_{\mu\nu}$ is the induced metric on ${\bf S}$, then the ``expansion''
 \eqn{Expansion}{
  \Theta \equiv h^{\mu\nu} \nabla_\mu \ell_\nu
 }
tells us whether deforming ${\bf S}$ in the $\ell^\mu$ direction makes it bigger or smaller.
 \begin{itemize}
  \item $\Theta < 0$ everywhere on ${\bf S}$ means that ${\bf S}$ is a trapped surface.
  \item $\Theta = 0$ everywhere on ${\bf S}$ means that ${\bf S}$ is a marginally trapped surface.
 \end{itemize}
The expansion of $n^\mu$ can be defined in a similar fashion, and, at least for simple choices of ${\bf S}$, it is automatically negative.

Penrose introduced a standard choice of marginally trapped surface---understood as the outer ``hull'' of a family of trapped surfaces---for shocks colliding in ${\bf R}^{3,1}$.  His choice is easily generalized to $AdS_5$.  Figure~\ref{3Dtrapped} shows a schematic depiction of this generalized Penrose construction.  The surface is the union of two halves, call them ${\bf S}_1$ and ${\bf S}_2$.  ${\bf S}_1$ is the surface specified by the equations
 \eqn{Sone}{
  u=0 \qquad\qquad v = -{L \over z} \Psi(x^1,x^2,z) \,,
 }
while ${\bf S}_2$ is specified by
 \eqn{Stwo}{
  v=0 \qquad\qquad u = -{L \over z} \Psi(x^1,x^2,z) \,.
 }
The function $\Psi$ has yet to be determined, so \eno{Sone} and \eno{Stwo} should be regarded at this point simply as ans\"atze.  Actually, ${\bf S}_1$ and ${\bf S}_2$ should be located just slightly forward in time from the trajectories $u=0$ and $v=0$ of the shocks.  Thus ${\bf S}_1$ ``feels'' the effects of the right-moving shock, and ${\bf S}_2$ feels the effects of the left-moving one.

A key point is that the cross-section of $AdS_5$ transverse to either of the trajectories of the massless particles is the hyperbolic space ${\bf H}_3$, parametrized by $x^1$, $x^2$, and $z$, with metric
 \eqn{HthreeMetric}{
  ds^2_{H_3} = {L^2 \over z^2} \left[ (dx^1)^2 + (dx^2)^2 + dz^2 \right]
 }
inherited from $AdS_5$.

With some work, and after using coordinate shifts like\footnote{$\theta(x)$ is the Heaviside step function, taking values $1$ when $x > 0$ and $0$ when $x < 0$.}
 \eqn{vShift}{
  v \to v + {L \over z} \Phi(x^1,x^2,z) \theta(u) \,,
 }
to get rid of distributional terms in the shock-wave metric, one can check the following claims:
 \begin{itemize}
  \item The marginal trapped surface equation $\Theta = 0$ on ${\bf S}_1$ and ${\bf S}_2$ boils down to
 \eqn{PsiPDE}{
  \left( \square_{H_3} - {3 \over L^2} \right) (\Psi-\Phi) = 0 \,,
 }
where $\square_{H_3}$ is the laplacian for the metric $ds_{H_3}^2$.
  \item Continuity of $\ell^\mu$ as one passes from ${\bf S}_1$ to ${\bf S}_2$ across the closed curve ${\bf C} = {\bf S}_1 \cap {\bf S}_2$ amounts to requiring
 \eqn{PsiBCs}{
  \Psi\bigg|_{\bf C} = 0 \qquad\hbox{and}\qquad
  (\partial\Psi)^2\bigg|_{\bf C} = 4 \,.
 }
Here $(\partial\Psi)^2 = g^{\mu\nu} \partial_\mu \Psi \partial_\nu \Psi$ where $g^{\mu\nu}$ is the metric on $H_3$.
  \item If we parametrize ${\bf S}_1$ by $(x^1,x^2,z)$, then the induced metric on ${\bf S}_1$ is exactly the metric we wrote for ${\bf H}_3$.  The same goes for ${\bf S}_2$.
 \end{itemize}

\subsection{Trapped surfaces respecting the $O(3)$ symmetry}
\label{OTHREE}

The reason we can handle the trapped surface in an analytical fashion is that it's highly symmetrical.  The purpose of this section is to explain the relevant symmetry, which we will put to greater use in the next lecture.  Given a point $x^1=x^2=0$, $z=L$ in ${\bf H}_3$---call this point $P$---the locus of points at a fixed geodesic distance away is a copy of $S^2$.  The $O(3)$ we're interested in is the one that acts by ordinary rotations on this $S^2$.  The quantity
 \eqn{qAgain}{
  q = {(x^1)^2 + (x^2)^2 + (z-L)^2 \over 4zL}
 }
which we encountered earlier is a simple function of the geodesic distance from $P$ to $(x^1,x^2,z)$.

$\square_{H_3}$ respects the $O(3)$ symmetry, as does $\Phi = \Phi(q)$, so it must be possible to solve the main PDE \eno{PsiPDE} with a function $\Psi(q)$.  To find the area of ${\bf S}_1$, we don't even need to know $\Psi(q)$ in detail: it suffices to know the value $q_{\rm C}$ where $\Psi(q_{\rm C}) = 0$, subject to the condition $(\partial\Psi)^2\bigg|_{\rm C} = 4$.  These boundary conditions also respect the $O(3)$ symmetry.

Given $q_{\rm C}$, we can compute the area $A_{\rm trapped}$ of the whole trapped surface ${\bf S}_1 \cup {\bf S}_2$ as twice the volume in ${\bf H}_3$ of the ball whose boundary is the sphere $S^2$ at ``radius'' $q_{\rm C}$.  Then $S_{\rm trapped}$ is computed from the formula
 \eqn{StrappedForm}{
  S_{\rm trapped} = {A_{\rm trapped} \over 4G_5} \,.
 }
With some work, one can derive a relation between $S_{\rm trapped}$ and the energy $E$ of one of the lightlike particles.  (It is assumed that the energies of the two massless particles are equal.  If this weren't true, an appropriate Lorentz boost would make it true.)  This relation is best expressed parametrically in terms of $q_{\rm C}$:
 \eqn{ESparam}{\eqalign{
  {EG_5 \over L^2} &= 2q_{\rm C}(1+q_{\rm C})(1+2q_{\rm C})  \cr
  {S_{\rm trapped} G_5 \over L^3} &= \pi (x_{\rm C} \sqrt{1+x_{\rm C}^2} - 
    \arcsinh x_{\rm C})
 }}
where
 \eqn{xcDef}{
  x_{\rm C} = 2 \sqrt{q_{\rm C}(1+q_{\rm C})} \,.
 }

The main result \eno{SBH} can be obtained by expanding the relations \eno{ESparam} at leading order in large $q_{\rm C}$ and $x_{\rm C}$.  So in a sense we're done.  But I've left out a lot, both in the derivation of the result and in its significance and possible generalizations.  I'll outline here a number of points which the interested reader could explore further.  Many of these points are well addressed in existing literature, for instance these works\cite{Gubser:2008pc,Gubser:2009sx,Chesler:2010bi} and references therein.  My main reason for providing such a long list of questions here is that these were approximately the questions I was asking myself at the stage of understanding where I had the main result \eno{SBH} more or less straight but was not fully confident of all the supporting details.
 \begin{itemize}
  \item How do you arrive at the shock wave metric \eno{dsTwo} with the specific functional form \eno{PhiForm}?
  \item How do you get $\langle T_{uu} \rangle = {2L^4 E \over \pi \left[ L^2 + (x^1)^2 + (x^2)^2 \right]^3} \delta(u)$ starting from the shock wave metric?
  \item How do you show that $\Theta < 0$ everywhere is the condition for a trapped surface?
  \item In what generality can you show that a marginally trapped surface (one with $\Theta=0$) is the outer hull of a family of trapped surfaces?
  \item How do you show that trapped surfaces have to be behind an event horizon?
  \item How do you derive \eno{PsiPDE} from $\Theta = 0$ and $(\partial\Psi)^2 \bigg|_{\rm C} = 4$ from continuity of $\ell^\mu$?
  \item How do you choose parameters $G_5$, $L$, and $E$ for a meaningful comparison to heavy ion physics?  How meaningful is this comparison?
  \item How do you estimate the entropy produced in a heavy ion collision starting from data?
  \item What other theoretical approaches are there for getting at this entropy?
  \item How does the main result, \eno{SBH}, generalize to other dimensions?
  \item Given that the dependence $S \propto E^{2/3}$ is somewhat too rapid as compared to data, are there sensible ways to change the $AdS_5$ calculation that would improve the match to data?
  \item Can one handle the case of off-center collisions, at least in some approximation?
  \item Can one at least approximately solve Einstein's equations in the future region, $u>0$ and $v>u$, and extract some useful information about thermalization, stopping, and the rapidity distribution of matter produced in the collision?
  \item What's the significance of the $O(3)$ symmetry in field theory terms?
 \end{itemize}

\section{Lecture 2: $O(3)$ symmetry and Bjorken flow}
\label{SYMMETRICAL}

Of all the questions I listed, the one that stuck with me the most was the last: Exactly what does the $O(3)$ symmetry do for you in the dual field theory?  Obviously it's crucial for the entire trapped surface story: it would be very difficult to find the function $\Psi$ if you had to deal with \eno{PsiPDE} as a true PDE in three variables.

Motivated by a discussion of RHIC-phenomenological hydrodynamics with U.~Heinz, I decided to look into how the $O(3)$ symmetry might constrain conformal relativistic hydrodynamics.  It turns out that there's a free lunch waiting to be eaten here: a rare treat!  To get at it, the first step is to understand Bjorken flow.\cite{Bjorken:1982qr}

\subsection{Bjorken flow}
\label{BJORKEN}

Consider a collision of highly relativistic heavy ions.  The causal future of the collision plane in ${\bf R}^{3,1}$ can be parametrized as
 \eqn{CausalFuture}{
  \begin{pmatrix} t \\ x^3 \\ x^1 \\ x^2 \end{pmatrix} = 
   \begin{pmatrix} \tau \cosh\eta \\ \tau \sinh\eta \\ x_\perp \cos\phi \\
     x_\perp \sin\phi \end{pmatrix} \,.
 }
The Minkowski metric can be expressed as
 \eqn{FlatMetric}{
  ds_{{\bf R}^{3,1}}^2 = -d\tau^2 + \tau^2 d\eta^2 + dx_\perp^2 + x_\perp^2 d\phi^2 \,.
 }
The variables $(\tau,\eta,x_\perp,\phi)$ make obvious two commuting isometries of ${\bf R}^{3,1}$: $SO(1,1)$ boosts, generated by $\partial / \partial\eta$, and $SO(2)$ rotations, generated by $\partial / \partial\phi$.

Part of Bjorken's setup is to assume that the post-thermalization dynamics of the QGP approximately respects both of these symmetries.  For $SO(2)$, that's pretty trivial: it just means that we're focusing on head-on collisions.  $SO(1,1)$ symmetry is {\it not at all} obvious, and it can only hold not too far from $\eta=0$.  Bjorken gave an argument\cite{Bjorken:1982qr} for $SO(1,1)$ symmetry based on how scattering of small $x$ partons leads to a locally thermalized medium.

Bjorken also assumed symmetry under the translations generated by $\partial/\partial x^1$ and $\partial/\partial x^2$.  This means that the finite size nucleus is replaced by an infinite sheet of matter filling the $x^1$-$x^2$ plane.  Conformal relativistic hydrodynamics constrained by the symmetries 
 \eqn{SymmetryFlat}{
  \left\{ {\partial \over \partial\eta}, {\partial \over \partial\phi}, {\partial \over \partial x^1}, {\partial \over \partial x^2} \right\}
 }
can be {\it solved exactly}.\footnote{It is not necessary to make the assumption of conformal symmetry at this point.  I do so for later convenience, and because the conformal equation of state $p = \epsilon/3$ is semi-realistic for highly energetic collisions.}  Here's how it's done.

The stress tensor takes the form
 \eqn{TmnForm}{
  T_{mn} = \epsilon u_m u_n + p (g_{mn} + u_m u_n) + \hbox{(viscous corrections)} \,,
 }
and conformal symmetry dictates $p = \epsilon/3$, so that $T^m{}_m = 0$.  The local four-velocity $u_m$ is constrained to have unit norm: $g^{mn} u_m u_n = -1$.  The symmetries listed in \eno{SymmetryFlat} imply that $u_m$ can only depend on $\tau$, and that $u_{x^1} = u_{x^2} = 0$.  An additional simplification which is reasonable in the center-of-mass frame of the collision of identical nuclei is to set $u_\eta = 0$.  So only $u_\tau$ is non-zero, and the unit norm constraint dictates that $u_\tau = -1$.  (The sign of $u_\tau$ is fixed by requiring that $u^m$ should be a future-directed vector, i.e.~$u^\tau > 0$.)

Straightforward calculations now show that the conservation equations, $\nabla^m T_{mn} = 0$, boil down in the inviscid case to
 \eqn{BjConserve}{
  {\partial \over \partial\tau} \left( \tau^{4/3} \epsilon \right) = 0 \,.
 }
So we conclude that
 \eqn{epsilonForm}{
  \epsilon = {\tilde\epsilon_0 \over \tau^{4/3}} \,,\qquad
   u^\tau = 1 \,, \quad\hbox{other $u^m=0$} \,,
 }
where $\tilde\epsilon_0$ is an integration constant.

\subsection{Massaging the symmetries}

Let's have another look at the {\it symmetries of Bjorken flow}, as listed in \eno{SymmetryFlat}.  Boost invariance commutes with the other three symmetries.  We can express
 \eqn{ddxOne}{\eqalign{
  {\partial \over \partial x^1} &= \cos\phi {\partial \over \partial x_\perp} - 
    {\sin\phi \over x_\perp} {\partial \over \partial\phi}  \cr
  {\partial \over \partial x^2} &= -\sin\phi {\partial \over \partial x_\perp} - 
    {\cos\phi \over x_\perp} {\partial \over \partial\phi} \,.
 }}
Starting from \eno{ddxOne}, it is easy to check the commutation relations
 \eqn{Commutators}{
  \left[ {\partial \over \partial\phi} , {\partial \over \partial x^1} \right] = 
    {\partial \over \partial x^2} \qquad\qquad
  \left[ {\partial \over \partial\phi} , {\partial \over \partial x^2} \right] = 
    -{\partial \over \partial x^1} \,.
 }
These relations should remind you of the $SO(3)$ commutators
 \eqn{SOthree}{
  [J^3,J^1] = i J^2 \qquad\qquad [J^3,J^2] = -i J^1 \,.
 }
But $\left[ {\partial \over \partial x^1} , {\partial \over \partial x^2} \right] = 0$, whereas $[J^1,J^2] = i J^3$.  So the generators $\left\{ {\partial \over \partial\phi}, {\partial \over \partial x^1}, {\partial \over \partial x^2} \right\}$ do {\it not} form the algebra $SO(3)$; instead they generate the group $ISO(2)$, which is a contraction of $SO(3)$, just as ${\bf R}^2$ is a large-radius limit of $S^2$.

I want to ``un-contract'' $ISO(2)$ back into $SO(3)$ while maintaining the property that all generators commute with $\partial/\partial\eta$.  The $SO(3)$ should be part of the group $SO(4,2)$ of conformal transformations of ${\bf R}^{3,1}$, which is also the group of isometries of $AdS_5$.  In fact, the $SO(3)$ I'm interested in is supposed to be the $SO(3)$ part of the $O(3)$ symmetry of ${\bf H}_3$ which figured prominently in section~\ref{OTHREE}.  If we insist that ${\partial/\partial\phi}$ should remain as one of the generators of $SO(3)$, then there's essentially only one possible deformation of ${\partial/\partial x^1}$ and ${\partial/\partial x^2}$ that will accomplish what we want:
 \eqn{ZetaUse}{
  {\partial \over \partial x^1} \to \zeta \qquad\qquad
  {\partial \over \partial x^2} \to \left[ {\partial \over \partial\phi}, \zeta
    \right] \,,
 }
where
 \eqn{ZetaDef}{\eqalign{
  \zeta &\equiv
    2q^2 \tau x_\perp \cos\phi {\partial \over \partial\tau} +
    (1+q^2\tau^2+q^2 x_\perp^2) \cos\phi {\partial \over \partial x_\perp}  \cr
   &\qquad\qquad{} - 
    {1+q^2\tau^2-q^2 x_\perp^2 \over x_\perp} \sin\phi {\partial \over \partial\phi}
 }}
and $q$ is a parameter with dimensions of momentum.  (Sorry, $q$ has nothing to do with distance on ${\bf H}_3$.)

Recall in the shock wave story that a single shock, $\langle T_{uu} \rangle = {2L^4 E \over \pi \left[ L^2 + (x^1)^2 + (x^2)^2 \right]^3} \delta(u)$, respects the $SO(3)$ symmetry in $AdS_5$ (but obviously, it does not respect $SO(1,1)$).  A slightly subtle analysis allows one one to identify $q = 1/L$.  Noting that $L$ is, in field theory terms, the $\langle T_{uu} \rangle$-weighted root-mean-square (rms) transverse radius of the boundary shockwave, we see that $1/q$ is essentially the transverse size of the colliding object.\footnote{$L$ is also the radius of $AdS_5$.  The astute reader may be wondering why this radius equals the rms transverse radius of the shockwave in field theory.  The answer is that I chose from the start to put the massless particles in $AdS_5$ at a depth $z=L$.  This simplifies some formulas, but it is inessential.  If I had put the massless particles at some other depth $z=z_*$, then in field theory, $L$ would get replaced by $z_*$.\cite{Gubser:2009sx}}

\subsection{Conformal isometries and hydrodynamics}

Bjorken was able to solve {\it completely} for $u_m$ just by demanding that $u_m$ respects the symmetries $SO(1,1) \times ISO(2)$, together with setting $u_\eta = 0$, which amounts to imposing an additional ${\bf Z}_2$ symmetry which acts as $\eta \to -\eta$.  In the previous section we saw how to replace $ISO(2)$ by $SO(3)$, which has just as many generators.  So we might ask, can we use the symmetry group $SO(1,1) \times SO(3) \times {\bf Z}_2$ to completely determine the velocity field $u_m$?

The answer is yes, but the details are a little tricky, and to explain them I'm going to have to remind you of how Lie derivatives work.  The defining relations are
 \eqn{LieDefs}{\eqalign{
  {\cal L}_\xi \phi &= \xi^n {\partial \over \partial x^n} \phi  \cr
  {\cal L}_\xi v^m &= \xi^n {\partial \over \partial x^n} v^m - 
    v^n {\partial \over \partial x^n} \xi^m  \cr
  {\cal L}_\xi \omega_m &= \xi^n {\partial \over \partial x^n} \omega_m + 
    \omega_n {\partial \over \partial x^m} \xi^n \,,
 }}
where $\xi^m$ and $v^m$ are vector fields, $\omega_n$ is a $1$-form, and $\phi$ is a scalar, all defined on ${\bf R}^{3,1}$, and all assumed to have appropriate smoothness properties.  ${\cal L}_\xi$ is linear, and it obeys Leibniz's Rule.  Crucially for purposes to come,
 \eqn{Lgmn}{
  {\cal L}_\xi g_{mn} = \nabla_m \xi_n + \nabla_n \xi_m = 0
 }
is the condition for $\xi^m$ to be an isometry of the metric $g_{mn}$.  What Bjorken did, in essence, to determine the four-velocity $u_m$, was to solve the equations 
 \eqn{FlatConstraints}{
  {\cal L}_\xi u_m = 0 \qquad\hbox{for}\qquad
   \xi \in \left\{ {\partial \over \partial\eta} , {\partial \over \partial\phi} , 
     {\partial \over \partial x^1} , {\partial \over \partial x^2} \right\} \,,
 }
with $u_\eta = 0$.

When we pass from $ISO(2)$ to $SO(3)$, the main complication is that the generators $\zeta$ and $\left[ \zeta, {\partial \over \partial\phi} \right]$ are not isometries of ${\bf R}^{3,1}$, but instead conformal isometries: for example,
 \eqn{ConformalIsometry}{
  {\cal L}_\zeta g_{mn} = {1 \over 2} (\nabla_\ell \zeta^\ell) g_{mn} \,.
 }
The result \eno{ConformalIsometry} is the infinitesimal statement of the fact that $\zeta^m$ generates conformal maps sending $g_{mn} \to \Omega^2 g_{mn}$, where the factor $\Omega$ depends on space and time.

We see from \eno{TmnForm} that the projection tensor $P_{mn} = g_{mn} + u_m u_n$ plays a key role in hydrodynamics.  Physically, this is the tensor which projects onto the spatial coordinates of the local rest frame of the fluid.  In order for $P_{mn}$ to transform nicely under conformal maps, we should demand that
 \eqn{LuRequire}{
  {\cal L}_\zeta u_m = {1 \over 4} (\nabla_\ell \zeta^\ell) u_m \,.
 }
Together with the constraints $u_\eta = 0$, ${\cal L}_{\partial \over \partial\eta} u_m = 0$, and ${\cal L}_{\partial \over \partial\phi} u_m = 0$, the equation \eno{LuRequire} is enough to determine $u_m$:
 \eqn{FoundU}{
  u^\tau = \cosh\kappa \qquad\qquad u^{x_\perp} = \sinh\kappa
 }
where
 \eqn{FoundKappa}{
  \kappa = \arctanh {2q^2 \tau x_\perp \over 1 + q^2 \tau^2 + q^2 x_\perp^2} \,,
 }
and $u^\eta = u^\phi = 0$.

For Bjorken flow, the next step would be to demand ${\cal L}_\xi \epsilon = 0$ for all the isometries $\xi$ (regarding $\epsilon$ as a scalar, so that ${\cal L}_\xi \epsilon = \xi^n {\partial \over \partial x^n} \epsilon$).  That would let us conclude that $\epsilon$ is a function only of $\tau$.  To obtain the explicit form for $\epsilon(\tau)$ that we listed in \eno{epsilonForm}, it's necessary to resort to the conservation equations $\nabla^m T_{mn} = 0$.

In the $SO(3)$-symmetric case, it still make sense to require ${\cal L}_\xi \epsilon = 0$ for $\xi = {\partial \over \partial\eta}$ and ${\partial \over \partial\phi}$: this just implies $\epsilon = \epsilon(\tau,x_\perp)$.  But ${\cal L}_\zeta \epsilon = 0$ might be the wrong equation, given that $\zeta$ is only a conformal isometry.  Instead let's try 
 \eqn{TryAlpha}{
  {\cal L}_\zeta \epsilon = -{\alpha \over 4} (\nabla_\ell \zeta^\ell) \epsilon \,,
 }
where $\alpha$ is a constant.\footnote{A considerably more systematic treat is possible here, based on the formalism introduced by Loganayagam\cite{Loganayagam:2008is}; but the way I am explaining the problem here is closer to what I actually did when I didn't know the answer.}

It's easy to show that the general solution to \eno{TryAlpha} is
 \eqn{FormEpsilon}{
  \epsilon = {\hat\epsilon(g) \over \tau^\alpha} \qquad\hbox{where}\qquad
   g = {1 - q^2 \tau^2 + q^2 x_\perp^2 \over 2q\tau} \,,
 }
and $\hat\epsilon(g)$ is an arbitrary function.  $g$ is essentially the only $SO(3)$-invariant combination of $\tau$ and $x_\perp$.  After a bit of work, one finds that the equations $\nabla^m T_{mn} = 0$ are consistent with one another iff $\alpha=4$, and that the general solution is
 \eqn{FoundEpsilonhat}{
  \hat\epsilon(g) = {\hat\epsilon_0 \over (1+g^2)^{4/3}} \,.
 }
Thus the $SO(3)$-invariant flow takes the final form
 \eqn{FinalFlowForm}{\eqalign{
  \epsilon &= {\hat\epsilon_0 \over \tau^{4/3}} {(2q)^{8/3} \over 
    \left[ 1 + 2q^2 (\tau^2 + x_\perp^2) + q^4 (\tau^2 - x_\perp^2)^2 \right]^{4/3}}  \cr
  v^\perp &\equiv {u^{x_\perp} \over u^\tau} = {2q^2 \tau x_\perp \over
    1 + q^2 \tau^2 + q^2 x_\perp^2} \,.
 }}

Let me close this lecture with a few thoughts on generalizations and relevance to phenomenology and gravity duals.
 \begin{itemize}
  \item One can add in viscous corrections and still get exact closed form expressions for $\epsilon$.  Of course, $u^m$ doesn't change: it is fixed by symmetry considerations alone.  Assorted other generalizations are possible, as explained in work that appeared after my TASI lectures were delivered.\cite{Gubser:2010ui}
  \item The transverse velocity $v^\perp$ is a quantity of considerable phenomenological interest (e.g. for single particle yields and Hanbury Brown-Twiss radii).  Having a symmetry argument that determines it is interesting, even if this symmetry is somewhat broken by real heavy ion collisions.
  \item I should stress that I do {\it not} claim that this deformation of Bjorken flow will describe $\langle T_{mn} \rangle$ in the dual of a point-sourced shock wave collision in $AdS_5$.  The $SO(3)$ symmetry is common to both situations, but in constructing a solution to hydro I have discarded all dynamical information from AdS/CFT and used instead the assumptions of that the flow has $SO(1,1)$ boost invariance and that hydrodynamics is valid.
 \end{itemize}

\section{Lecture 3: $p$-wave holographic superconductors}
\label{PWAVE}

\subsection{Overview of the main results}

\begin{figure}
\centerline{\psfig{file=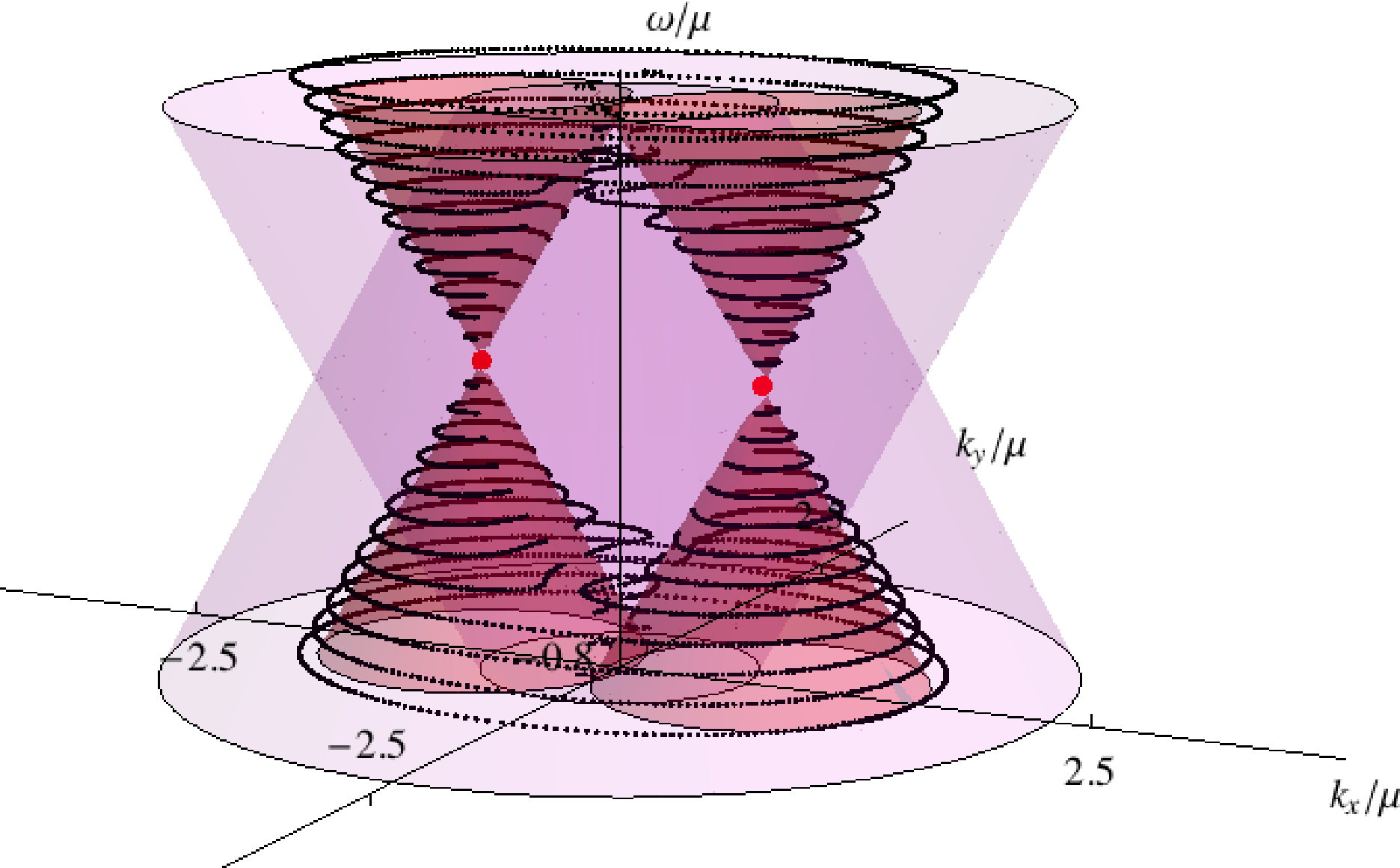,width=4in}}
\caption{The spectral structure of the two-point function $G^\Psi$ of a fermion operator in a $p$-wave holographic superconductor at zero temperature.  The dark cones are the Dirac cones whose apexes are the locations of gapless fermion excitations.  The black circles around the Dirac cones are the locations of fermion normal modes.  From the original literature.\cite{Gubser:2010dm}}\label{3dconeq3}
\end{figure}

In $p$-wave holographic superconductors,\cite{Gubser:2008zu,Gubser:2008wv,Roberts:2008ns} the Fermi surface degenerates to a pair of points, above each of which a Dirac cone rises, enclosing a continuum of fermion modes.  There are also discrete fermion normal modes slightly outside the Dirac cones.  See figure~\ref{3dconeq3}.  This structure is in contrast to the normal state, where the Fermi surface is a circle (when the field theory is in $2+1$ dimensions), and a non-trivial power law governs the response at small but non-zero frequencies.\cite{Liu:2009dm}

The underlying lagrangian on which all the calculations on this topic will be based is
 \eqn{EYM}{
  {\cal L} = R + {6 \over L^2} - {1 \over 2} \tr F_{\mu\nu}^2 - 
    i \Psi \Gamma^\mu D_\mu \Psi
 }
in bulk spacetime dimension $D=4$, where
 \eqn{FmnDef}{
  F_{\mu\nu} = \partial_\mu A_\nu - \partial_\nu A_\mu - 
    i g_{\rm YM} [A_\mu, A_\nu]
 }
and $A_\mu = A_\mu^a \tau^a$.  The matrices $\tau^a = {1 \over 2} \sigma^a$, with $a=1,2,3$, are the generators of $SU(2)$.  $\Psi$ is a doublet of $SU(2)$, and
 \eqn{DmDef}{
  D_\mu \Psi = (\nabla_\mu - i g_{\rm YM} A_\mu^a \tau^a) \Psi \,,
 }
where $\nabla_\mu$ includes the spin connection.

The simplest solution to the equations of motion is $AdS_4$:
 \eqn{AdSfour}{
  ds^2 = {L^2 \over z^2} \left[ -dt^2 + (dx^1)^2 + (dx^2)^2 + dz^2 \right] \,,
 }
with $F_{\mu\nu} = 0$.  The $T \to 0$ limit of the $p$-wave holographic superconductor is an $AdS_4$ to $AdS_4$ domain wall\cite{Basu:2009vv} (c.f.~the $T \to 0$ limit of RNAdS, which interpolates between $AdS_4$ and $AdS_2 \times {\bf R}^2$).  The gauge field interpolates between two flat gauge connections: $A_\mu^{\rm UV}$ near the boundary, and $A_\mu^{\rm IR}$ in the deep infrared.

The main results on fermion two-point functions at zero temperature\cite{Gubser:2010dm} can be understood starting from the gauge-covariant wave-vector:
 \eqn{KmDef}{
  K_m \equiv k_m - g_{\rm YM} A_m = k_m {\bf 1} - g_{\rm YM} A_m^a \tau^a \,,
 }
where ${\bf 1}$ is the $2 \times 2$ identity matrix.  We also define
 \eqn{KIRUV}{
  K_m^{\rm IR} = k_m - g_{\rm YM} A_m^{\rm IR} \qquad\qquad
  K_m^{\rm UV} = k_m - g_{\rm YM} A_m^{\rm UV} \,.
 }
In \eno{KmDef}, \eno{KIRUV}, and below, the index $m$ runs over boundary directions, i.e.~$m=0,1,2$.  Also, we require $A_z=0$ throughout: this is a gauge choice.  If ${\cal O}_\Psi$ is the operator in field theory dual to the fermion $\Psi$, then
 \eqn{Ofct}{
  \langle {\cal O}_\Psi^\dagger {\cal O}_\Psi \rangle \sim
    G_{\rm sudden}^\Psi(k) \equiv -i (\gamma^m q_m)^{-1}
      (q + \gamma^m \gamma^n q_{mn}) \gamma^t \,,
 }
where
 \eqn{qDefs}{\eqalign{
  q &= K_{\rm IR} \cosh(z_* K_{\rm UV})  \cr
  q_m &= -i \left[ K_m^{\rm IR} \cosh(z_* K_{\rm UV}) + K_{\rm IR} K_m^{\rm UV}
     {\sinh(z_* K_{\rm UV}) \over K_{\rm UV}} \right]  \cr
  q_{mn} &= K_m^{\rm IR} K_n^{\rm UV} {\sinh(z_* K_{\rm UV}) \over K_{\rm UV}}
 }}
and
 \eqn{KnormDefs}{
  K_{\rm IR} = \sqrt{\eta^{mn} K_m^{\rm IR} K_n^{\rm IR}} \qquad
  K_{\rm UV} = \sqrt{\eta^{mn} K_m^{\rm UV} K_n^{\rm UV}} \,.
 }
As I will explain, $z_*$ is a length scale characteristic of the bosonic background interpolating between $AdS_4$ and $AdS_4$.  $G_{\rm sudden}^\Psi(k)$ is the expression obtained for the fermion correlator when this background is treated in a thin wall approximation.  This approximation is not controlled in the sense of being approached as one dials a parameter of the lagrangian to an extreme value.  However, it captures the qualitative features of the fermion two-point function, which is otherwise accessible only through numerics.

The continuous part of the spectral weight of $G_{\rm sudden}^\Psi(k)$ arises precisely where $K_{\rm IR}$ has a branch cut---i.e.~where $K_m^{\rm IR}$ is timelike---because otherwise, $G_{\rm sudden}^\Psi(k)$ is a rational function of the $k_m$.  $K_m^{\rm IR}$ is timelike inside the aforementioned Dirac cones.

The motivations for working out the fermion response to $p$-wave holographic superconductors were numerous:
 \begin{itemize}
  \item We knew about holographic superconductors, both $s$-wave and $p$-wave.
  \item We knew about fermion correlators in the normal state.
  \item To have some chance at successful comparison to ARPES, where Dirac cones above isolated points on the Fermi surface are observed, we knew we needed non-$s$-wave dynamics.
  \item The lagrangian we chose is almost completely determined at the two-derivative level by its symmetries: basically it's QCD with $N_c=2$ and $N_f=1$ (with lagrangian $-{1 \over 2} \tr F^2 - i \bar\Psi \slashed{D} \Psi)$ coupled to gravity with a negative cosmological constant (with lagrangian $R + {6 \over L^2}$).
  \item It's easy to get lagrangians similar to \eno{EYM} out of string/M-theory low-energy effective actions.
  \item The $AdS_4$ to $AdS_4$ domain wall structure had recently been explained.\cite{Basu:2009vv}
 \end{itemize}

\subsection{Generalities on holographic superconductors}
\label{SUPERCONDUCTORS}

The main macroscopic features of superconductors are a consequence of the spontaneous breaking of $U(1)_{\rm EM}$, at finite $T$ and finite chemical potential $\mu$ for charge carriers.  In much of the theory of superconductivity (including the classic Bardeen-Cooper-Schrieffer theory of low-temperature $s$-wave superconductors), $U(1)_{\rm EM}$ is treated as a global symmetry for purposes of calculations of the gap, the condensate, and other properties; it later can be weakly gauged.  In this spirit, consider a field theory on ${\bf R}^{2,1}$ with a global $U(1)$ symmetry and an $AdS_4$ dual.  The field content of the dual gravity theory must contain an abelian gauge field, dual to the conserved current $J^\mu$ in the boundary theory.  Thus, on the gravity side, we should consider
 \eqn{Lgravity}{
  {\cal L} = R + {6 \over L^2} - {1 \over 4} F_{\mu\nu}^2 + \hbox{(matter fields)} \,,
 }
where the matter fields can be charged or uncharged.

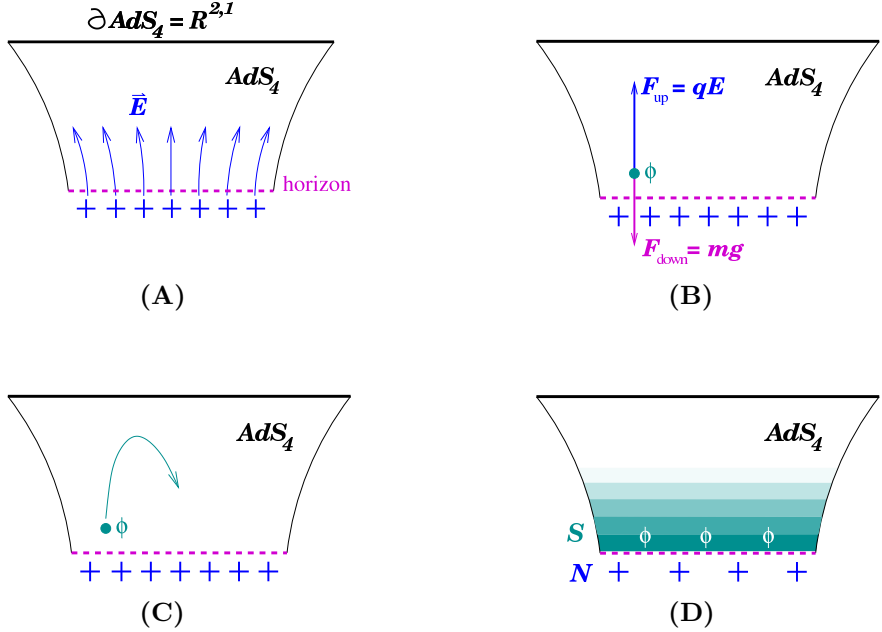
\begin{figure}
\begin{picture}(0,250)(0,-30)
\put(0,147){\psfig{file=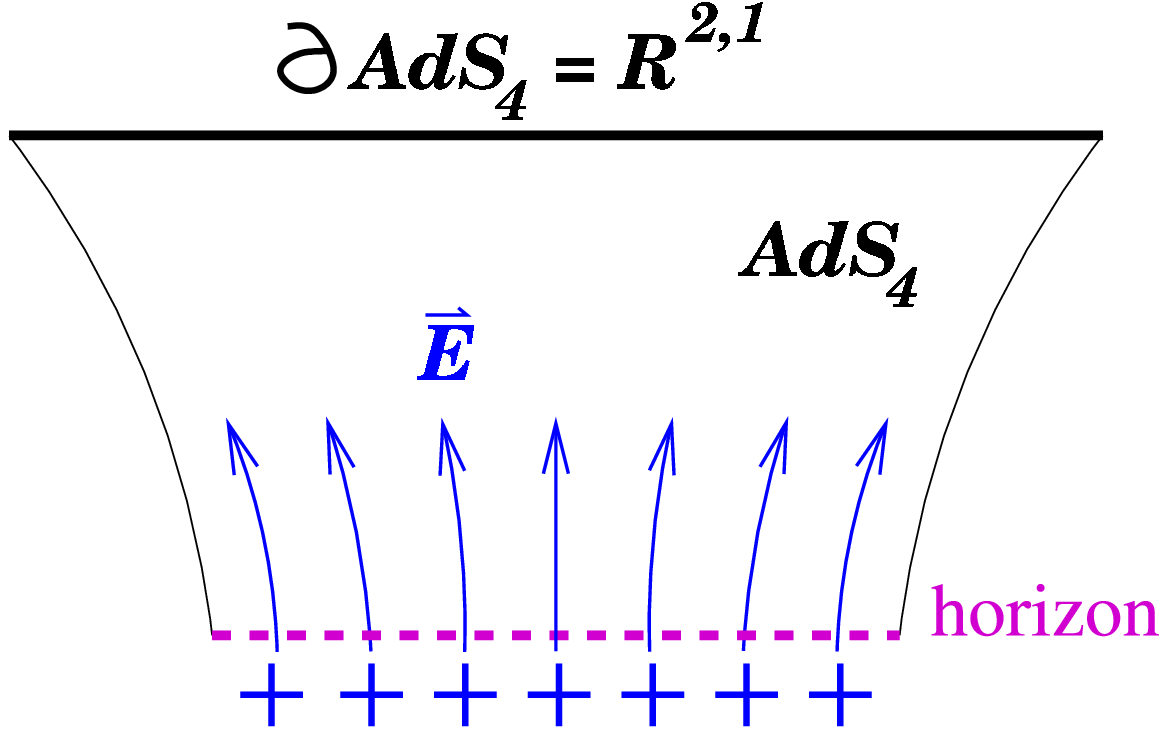,width=1.8in}}
\put(50,105){\bf (A)}
\put(200,120){\psfig{file=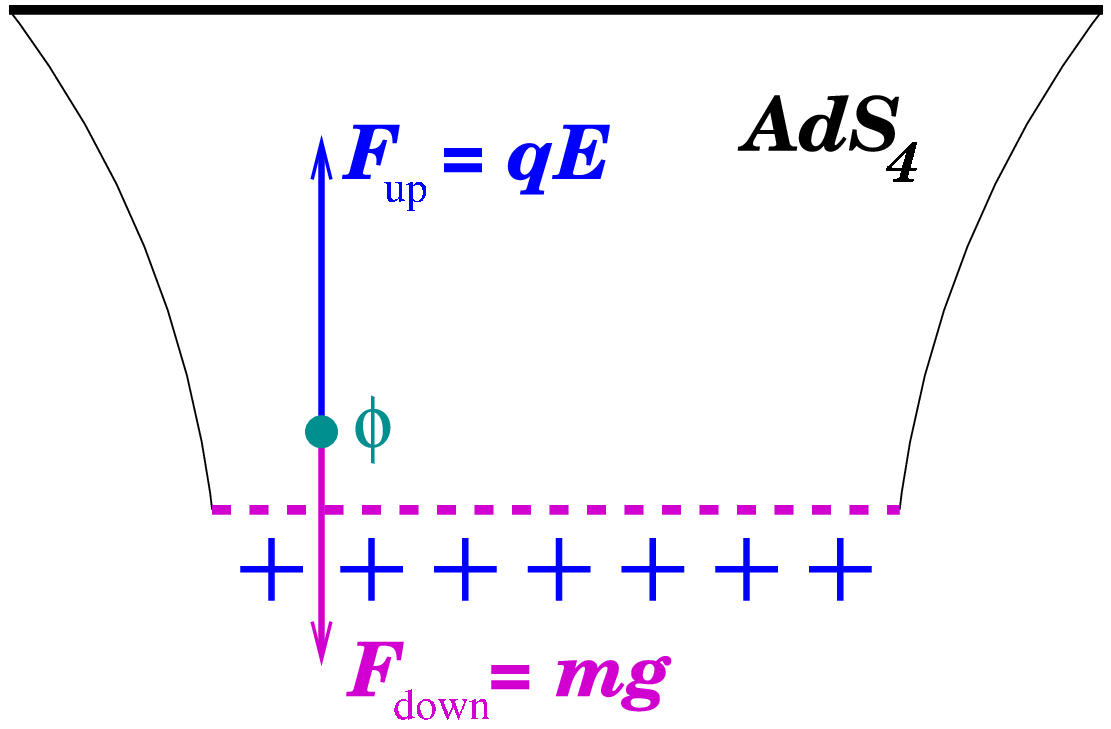,width=1.8in}}
\put(250,105){\bf (B)}
\put(0,10){\psfig{file=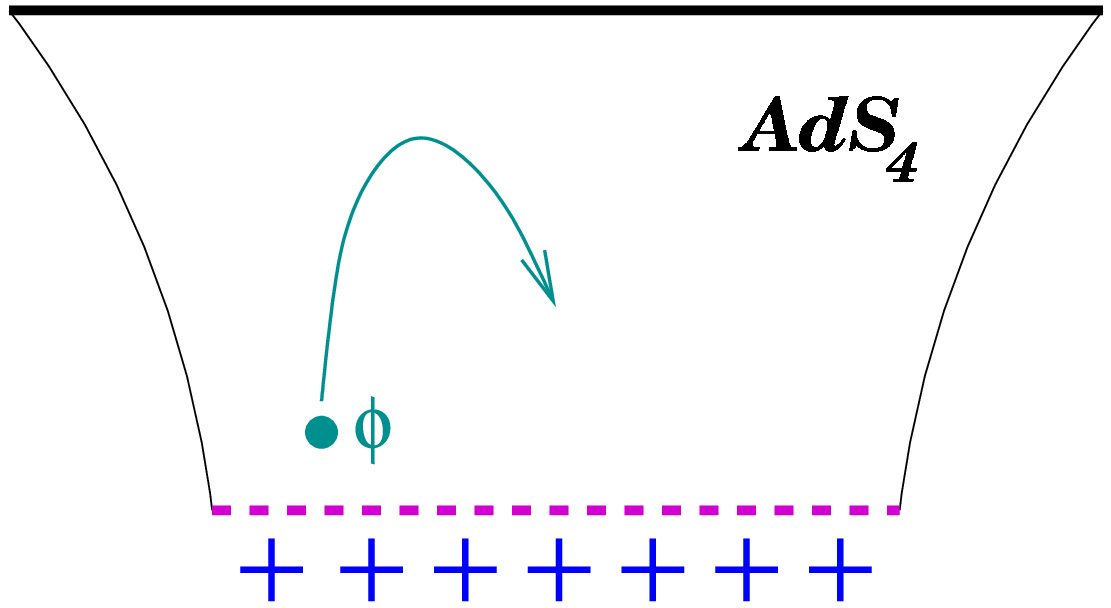,width=1.8in}}
\put(50,-15){\bf (C)}
\put(200,0){\psfig{file=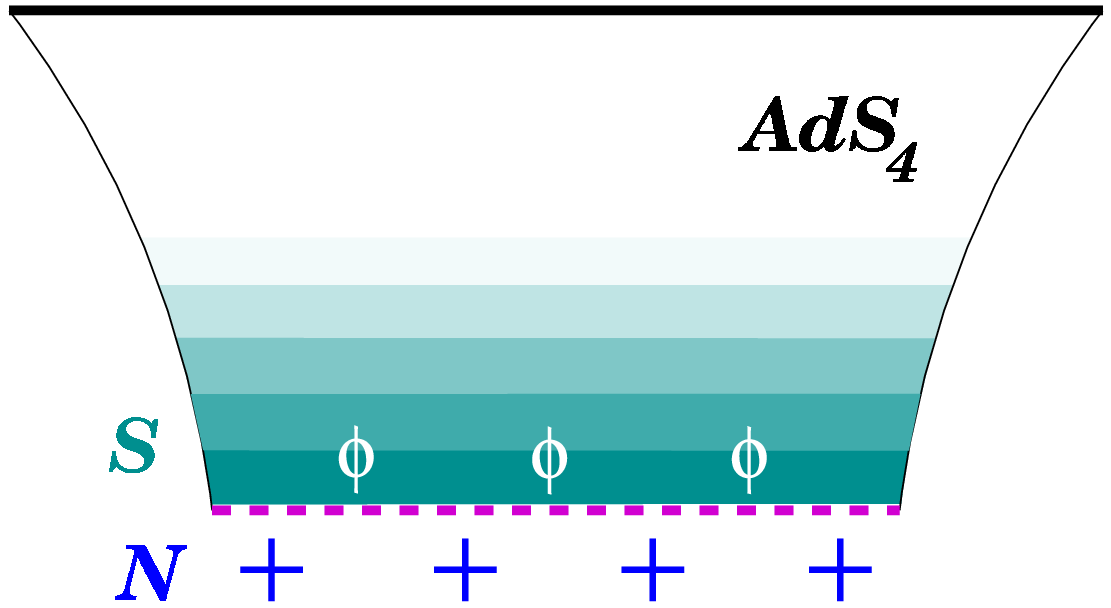,width=1.8in}}
\put(250,-15){\bf (D)}
\end{picture}
\caption{A qualitative account of holographic superconductors, in pictures.  (A) The normal state is described in terms of the RNAdS solution, with boundary ${\bf R}^{2,1}$.  (B) The upward electrostatic force on a charged quantum of a scalar field $\phi$ can be greater than the downward gravitational pull.  (C) The $AdS_4$ asymptotics prevent particles from escaping arbitrarily far from the horizon.  (D) The quanta of $\phi$ instead condense just outside the horizon.}\label{SCcartoon}
\end{figure}

First let's consider the $s$-wave case.\cite{Gubser:2008px,Hartnoll:2008vx}  The matter fields include
 \eqn{Lswave}{
  {\cal L}_\phi = -\left| (\partial_\mu - iq A_\mu)\phi \right|^2 - m^2 |\phi|^2 + 
    \ldots \,,
 }
where $\phi$ is a complex scalar.  In the normal state, $\phi=0$ because only vanishing $\phi$ is preserved by $U(1)$ rotations.  The simplest solution with non-zero gauge field is then Reissner-Nordstrom $AdS_4$ (RNAdS).  I will not need to consider the detailed form of RNAdS.  The qualitative features shown in figure~\ref{SCcartoon}A are enough.  In particular, the electric field is
 \eqn{Ez}{
  E_z = F_{0z} = -\partial_z \Phi
 }
where $A_0 = \Phi$.  The chemical potential is the amount of energy it takes to push a unit of charge from the boundary into the horizon:
 \eqn{FoundMu}{
  \mu = \Phi_{\rm bdy} - \Phi_{\rm horizon} \,.
 }
But $\Phi_{\rm horizon}$ must be set to $0$ in order for $A = \Phi dt$ to be well-defined at the horizon.

The key question is: Will the scalar condensed outside the horizon?  Heuristically, the answer is YES, provided $q \neq 0$, and $m$ is not too big, and $T$ is sufficiently small.  A naive way to reason this out is to consider the balance of forces on a test particle slightly above the horizon, as illustrated in figure~\ref{SCcartoon}B.  The electrostatic force $F_{\rm up} = qE$ overcomes the gravitational force $F_{\rm down} = mg$ provided $qE > mg$.  Then quanta of $\phi$ want to jump out of the black hole (see figure~\ref{SCcartoon}C).  But because of the infinite blueshift of the $AdS_4$ metric near the boundary, nothing can escape.  So the simplest endpoint for the dynamics is for the charged bosonic field to condense near the horizon, as in figure~\ref{SCcartoon}D.  Now, the surface gravity of the horizon is related to the Hawking temperature by $g = 2\pi T$, so we expect a condensate $\phi \neq 0$ for $T$ less than some critical value $T_c$.

The $p$-wave case\cite{Gubser:2008zu,Gubser:2008wv,Roberts:2008ns} is a variant of the $s$-wave story: Instead of using a complex scalar $\phi$ as the charged matter field, we promote $F_{\mu\nu}$ to an $SU(2)$ field strength.  If the original $U(1)$ is associated with the $\tau^3$ part of $SU(2)$, then $A_\mu^\pm \equiv A_\mu^1 \pm i A_\mu^2$ are fields with charges $q = \pm g_{\rm YM}$: this is just about how $W^\pm$ bosons arise, except that here we have no Higgs field and no $U(1)_Y$ hypercharge gauge group.  As mentioned earlier, work of Basu and collaborators\cite{Basu:2009vv} demonstrated that in the zero-temperature limit, $p$-wave holographic superconductors take the form of $AdS_4$-to-$AdS_4$ domain walls.  In the next section, I will explain a simplified version of this construction in a limit where the gauge field doesn't back-react on the metric.

\subsection{Domain wall backgrounds in the probe limit}

In the limit $g_{\rm YM} \to \infty$, the gauge field doesn't back-react on the geometry.  To see this, define
 \eqn{HatAF}{
  \hat{A}_\mu = A_\mu / g_{\rm YM} \qquad\qquad 
  \hat{F}_{\mu\nu} = \partial_\mu \hat{A}_\nu - \partial_\nu \hat{A}_\mu -
     i [\hat{A}_\mu, \hat{A}_\nu] \,.
 }
Then the bosonic lagrangian takes the form
 \eqn{Lprobe}{
  {\cal L} = R + {6 \over L^2} - {1 \over 2 g_{\rm YM}^2} \tr \hat{F}_{\mu\nu}^2 \,.
 }
The $1/g_{\rm YM}^2$ suppression means we can solve the equations of motion of ${\cal L}_{\rm grav} = R + {6 \over L^2}$ first to get (at zero temperature) $AdS_4$, as in \eno{AdSfour}; then we can solve the classical Yang-Mills equation in this background.  $AdS_4$ is conformal to the $z>0$ half of Minkowski space ${\bf R}^{3,1}$, and the classical Yang-Mills equations are conformally invariant.  So we can solve them on ${\bf R}^{3,1}$ instead of on $AdS_4$, starting with the ansatz
 \eqn{Aansatz}{
  \hat{A} = \Phi \tau^3 dt + W \tau^1 dx^1 \,.
 }
The $\Phi \tau^3 dt$ term in \eno{Aansatz} is needed in order to describe the $U(1)$ chemical potential.  The $W \tau^1 dx^1$ term is the simplest expression that spontaneously breaks the $U(1)$ generated by $\tau^3$.  Both $\Phi$ and $W$ are required to be functions only of $z$, and one can easily demonstrate that the Yang-Mills equations boil down to
 \eqn{dPhiW}{
  {d^2 \Phi \over dz^2} = W^2 \Phi \qquad\qquad
   {d^2 W \over dz^2} = -\Phi^2 W \,.
 }
Appropriate boundary conditions are
 \eqn{BCs}{\seqalign{\span\TL & \span\TR &\qquad\span\TT}{
  \Phi &\to 0 \,,\ \ W \to W_{\rm IR} & as $z \to \infty$ (the infrared)  \cr
  \Phi &\to \mu \,,\ \ W \to 0 & as $z \to 0$ (the ultraviolet)
 }}
Requiring $W \to 0$ in the ultraviolet is the condition that the symmetry breaking must be spontaneous: we are deforming the CFT lagrangian only by the $U(1)$-symmetric term $\mu J_0^3$, where $J_0^3$ is the charge density for the $\tau^3$ part of $SU(2)$.

\begin{figure}
\centerline{\psfig{file=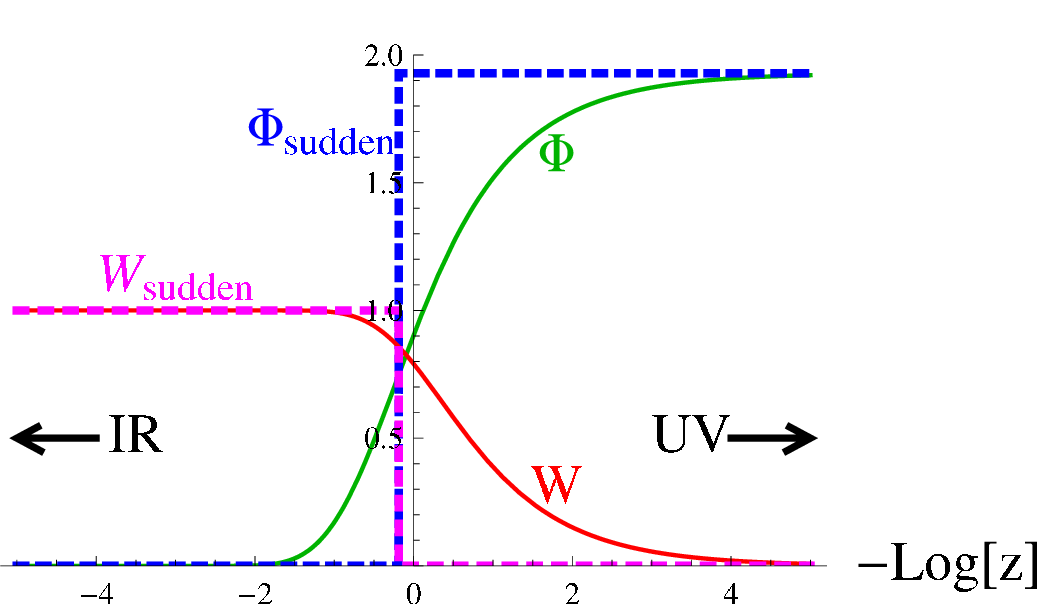,width=3in}}
\caption{The solution to \eno{dPhiW} with boundary conditions \eno{BCs}.  The step functions $\Phi_{\rm sudden}$ and $W_{\rm sudden}$, plotted as dashed lines, provide the thin wall, or sudden, approximation to this solution.}\label{SuddenApprox}
\end{figure}

The solution to \eno{dPhiW} with the boundary conditions \eno{BCs} is essentially unique, and it is shown in figure~\ref{SuddenApprox}.  Soon we will want to make a further approximation: replace $\Phi$ and $W$ by step functions:
 \eqn{WPhiSudden}{
  W_{\rm sudden}(z) = W_{\rm IR} \, \theta(z_*-z) \qquad\qquad
  \Phi_{\rm sudden}(z) = \mu \, \theta(z-z_*) \,,
 }
where $z_*$ is defined so that
 \eqn{zstarCondition}{
  \int_0^\infty dz \, \Phi_{\rm sudden}(z) = \int_0^\infty dz \, \Phi(z) \,.
 }

\section{Lecture 4: Fermion correlators and the sudden approximation}
\label{FERMIONS}

\subsection{Extracting the fermion two-point function}

We saw in the previous lecture that constructing a $p$-wave holographic superconductor at zero temperature reduces in the large $g_{\rm YM}$ limit to finding a domain wall solution to the classical Yang-Mills equations in the $z>0$ half of flat four-dimensional Minkowski space.  This is charming, because most holographic superconductors are governed by more complicated equations.  What made things work is the invariance of the Yang-Mills equations under conformal transformations.  The massless Dirac equation is also essentially invariant under conformal transformations: Defining
 \eqn{UpperLowerPsi}{
  \psi = \left( {L \over z} \right)^{3/2} \Psi \,,
 }
where $\psi$ is regarded as a fermion on ${\bf R}^{3,1}$ and $\Psi$ is the original fermion on $AdS_4$, one finds that
 \eqn{AdSDirac}{
  \Gamma^\mu (\nabla_\mu - i \hat{A}_\mu) \Psi = 0
 }
on $AdS_4$ is equivalent to
 \eqn{FlatDirac}{
  \Gamma^\mu (\partial_\mu - i \hat{A}_\mu) \psi = 0
 }
on the $z>0$ part of ${\bf R}^{3,1}$.

The obvious ansatz for solving \eno{FlatDirac} is
 \eqn{psiAnsatz}{
  \psi(x^m,z) = e^{i k_m x^m} \hat\psi(z) \,.
 }
Recall that we defined a covariant wave-vector as
 \eqn{KmDefAgain}{
  K_m = k_m - \hat{A}_m \,.
 }
Plugging \eno{psiAnsatz} into \eno{FlatDirac} and using \eno{KmDefAgain} leads directly to $(\Gamma^m i K_m + \Gamma^z \partial_z) \psi = 0$, and recalling that $(\Gamma^z)^2 = 1$, we see that
 \eqn{dzPsi}{
  (\partial_z + i \Gamma^z \Gamma^m K_m) \hat\psi = 0 \,.
 }
Formally, the solution to this equation is
 \eqn{PsiHatFormal}{
  \hat\psi = P \left\{ e^{-i\int_0^z dz' \, \Gamma^z \Gamma^m K_m(z')} \right\}
    \hat\psi(0) \,.
 }
The solution we want for computing Green's functions has asymptotic behavior
 \eqn{psiHatBehave}{
  \hat\psi \propto e^{-K_{\rm IR} z} u \qquad\hbox{for large $z$,}
 }
where $u$ is a constant spinor.  The reason that the solution \eno{psiHatBehave} is preferred over the solution proportional to $e^{+K_{\rm IR} z}$ is that only the solution \eno{psiHatBehave} has the property that $\hat\psi \to 0$ as $z \to \infty$ when $K_m^{\rm IR}$ is spacelike.\footnote{It is slightly subtle to say what we mean in describing $K_m^{\rm IR}$ as spacelike, because each of the $K_m^{\rm IR}$ is a matrix.  Fortunately, they are commuting matrices, so one can find simultaneous eigenspaces.  On any one of these simultaneous eigenspaces, the $K_m^{\rm IR}$ act as numbers, and it is these numbers which are assumed to fill out a spacelike three-vector.}  When $K_m^{\rm IR}$ is not spacelike, then a $+i\epsilon$ prescription can be used to formulate an appropriate pole-passing description that fully defines the Green's function.  For the retarded Green's function, this $+i\epsilon$ prescription can be found by demanding that \eno{psiHatBehave} is infalling at the horizon when $K_m^{\rm IR}$ is not spacelike.

Noting the trivial identity $(\partial_z + K_{\rm IR}) e^{-K_{\rm IR} z} u = 0$, we see that the condition on $\hat\psi(z)$ is
 \eqn{PsiHatOnce}{
  \lim_{z\to 0} (K_{\rm IR} - i \Gamma^z \Gamma^m K_m) \hat\psi(z) = 0 \,,
 }
which is equivalent to
 \eqn{PsiHatTwice}{
  (K_{\rm IR} - i \Gamma^z \Gamma^m K_m^{\rm IR}) P \left\{ 
    e^{-i\int_0^\infty dz \, \Gamma^z \Gamma^m K_m(z)} \right\} \hat\psi(0) = 0 \,.
 }
Formally, \eno{PsiHatTwice} takes the form ${\bf P} \hat\psi(0) = 0$ for a matrix ${\bf P}$ that has four-valued Dirac and two-valued $SU(2)$ indices.

The usual basis for $\Gamma^\mu$ in this type of calculation is
 \eqn{GammaBasis}{
  \Gamma^m = \begin{pmatrix} 0 & \gamma^m \\ \gamma^m & 0 \end{pmatrix} \qquad\qquad
  \Gamma^z = \begin{pmatrix} -1 & 0 \\ 0 & 1 \end{pmatrix} \,,
 }
where
 \eqn{gammas}{
  \gamma^t = i\sigma_2 \qquad
  \gamma^1 = \sigma_1 \qquad
  \gamma^2 = \sigma_3 \,,
 }
so that $\{ \gamma^m, \gamma^n \} = 2\eta^{mn} = 2\diag\{ -1,1,1 \}$.  If in this basis we express
 \eqn{Pblocks}{
  {\bf P} = \begin{pmatrix} P_{++} & P_{+-} \\ P_{-+} & P_{--} \end{pmatrix} \,,
 }
then
 \eqn{GkForm}{
  G^\Psi(k) = i P_{+-}^{-1} P_{++} \gamma^t = i P_{--}^{-1} P_{-+} \gamma^t \,.
 }
A justification of \eno{GkForm} can be found in the original literature.\cite{Gubser:2010dm}

\subsection{Simplifications based on the sudden approximation}

Because the $K_m(z)$'s don't commute at different values of $z$, we need something extra to make further progress with analytical methods.  So let's use the sudden approximation:
 \eqn{psiHatSudden}{
  \hat\psi_{\rm sudden}(z) = \left\{ \seqalign{\span\TL\quad & \span\TT}{
    e^{-iz \Gamma^z \Gamma^m K_m^{\rm UV}} \hat\psi(0) & for $0 < z < z_*$  \cr
    e^{-i(z-z_*) \Gamma^z \Gamma^m K_m^{\rm IR}} e^{-i z_* \Gamma^z \Gamma^m K_m^{\rm UV}}
      \hat\psi(0) & for $z>z_*$ \,,} \right.
 }
where we take advantage of the fact that the $K_m^{\rm IR}$ commute with one another, as do the $K_m^{\rm UV}$.  Now,
 \eqn{KIRone}{
  (K_{\rm IR} - i \Gamma^z \Gamma^m K_m^{\rm IR}) \hat\psi_{\rm sudden}(z) = 0 \qquad
   \hbox{for all $z \geq z_*$}
 }
if and only if
 \eqn{KIRtwo}{
  (K_{\rm IR} - i \Gamma^z \Gamma^m K_m^{\rm IR}) e^{-i z_* \Gamma^m K_m^{\rm UV}}
    \hat\psi(0) = 0 \,.
 }
Thus to compute $G^\Psi_{\rm sudden}(k)$, which is the sudden approximation to the fermion two-point function $G^\Psi(k)$, we can use
 \eqn{Pvalue}{
  {\bf P} = (K_{\rm IR} - i \Gamma^z \Gamma^m K^{\rm IR}_m) e^{-i z_* \Gamma^m K_m^{\rm UV}}
          = q + \Gamma^z \Gamma^m q_m + \Gamma^m \Gamma^n q_{mn} \,,
 }
where $q$, $q_m$, and $q_{mn}$ are given in \eno{qDefs}.  The point of the decomposition \eno{Pvalue} is to decouple the spinor structure and the $SU(2)$ structure.  $q$, $q_m$, and $q_{mn}$ are all spinor singlets and $SU(2)$ adjoints.

Just a bit more work with $\gamma^m$ matrices allows us to demonstrate the result I claimed earlier as \eno{Ofct} and reproduce here for convenience:
 \eqn{Gsudden}{
  G_{\rm sudden}^\Psi(k) = -i (\gamma^m q_m)^{-1} (q + \gamma^m \gamma^n q_{mn}) \gamma^t 
   \,.
 }
Note that $q$, $q_m$, and $q_{mn}$ are {\it analytic} functions of $K_m^{\rm UV}$, because $\cosh(z_* K_{\rm UV})$ and ${\sinh(z_* K_{\rm UV}) \over K_{\rm UV}}$ are really functions of $K_{\rm UV}^2 \equiv \eta^{mn} K_m^{\rm UV} K_n^{\rm UV}$.  The {\it spectral weight} of $G_{\rm sudden}^\Psi(k)$ arises from poles and branch cuts of $G^\Psi_{\rm sudden}$ as a function of the $k_m$.  As I already reviewed, branch cuts can only come from the square root in $K_{\rm IR} \equiv \sqrt{\eta^{mn} K_m^{\rm IR} K_n^{\rm IR}}$, and therefore occur precisely when $K_m^{\rm IR}$ is timelike.  In such a case, $K_{\rm IR}$ is imaginary and $e^{-K_{\rm IR} z}$ is {\it oscillatory}: infalling if we're computing a retarded Green's function.

Let me now develop the analytic structure of $G^\Psi_{\rm sudden}(k)$ in more explicit detail.  The $x^1$ component of the gauge field is $\hat{A}_1^{\rm IR} = W_{\rm IR} \tau^1$, whose eigenvalues are $\pm k_*$ where
 \eqn{kstarDef}{
  k_* = {W_{\rm IR} \over 2} \,.
 }
$K_m^{\rm IR}$ therefore has eigenvalues $k_m - k_{\lambda,m}$, where $\lambda = \pm 1/2$ and
 \eqn{klambdaDef}{
  k_{\pm {1 \over 2}, m} = (0,\pm k_*,0) \,.
 }
What really matters is if $k_m - k_{\lambda,m}$ is timelike: If it is, then there's a branch cut in the $\lambda$ eigenspace of $K_{\rm IR}$, and hence in $G_{\rm sudden}^\Psi$.  This is what leads to the Dirac cones, which rise above the two isolated points $k_{\lambda,m}$ in phase space.

Besides square root branch cuts, there is {\it another} way to get spectral weight: $(\gamma^m q_m)^{-1}$ might have a pole.  Let's inquire when this could happen.  If $K_{\rm IR}^m$ has a timelike part, then $K_{\rm IR}$ has an anti-hermitian part, and it would be non-generic for
 \eqn{gqm}{
  \gamma^m q_m = -i \left[ \gamma^m K_m^{\rm IR} \cosh(z_* K_{\rm UV}) + 
    K_{\rm IR} \gamma^m K_m^{\rm UV} {\sinh(z_* K_{\rm UV}) \over K_{\rm UV}} \right]
 }
to be non-invertible.  On the other hand, if $K_{\rm IR}$ is hermitian (which happens when $K_m^{\rm IR}$ is spacelike), then adjusting one parameter (e.g.~$\omega$ with $k^1$ and $k^2$ held fixed) will make $\det \gamma^m q_m$ vanish.  The conclusion is that there can generically be a pole in $G_{\rm sudden}^\Psi(k)$ outside the Dirac cones, but not inside.

I would hasten to point out that the argument of the previous paragraph is not airtight---unlike the analysis of where branch cuts appear, which is pretty obviously the complete story.  In order to probe the question further,\cite{Gubser:2010dm} we looked numerically for poles in $G^\Psi(k)$, and we indeed found just one continuous locus of zeros outside the Dirac cones.  This locus eventually intersects the edge of the Dirac cones, as if the zeros were trying to get into the region where there are branch cuts.  When this happens, it corresponds to a stable excitation (corresponding to a pole in $G^\Psi(k)$ at real values of $k^m$) becoming a sharp but finite-width resonance.

\subsection{Further developments}

As with the previous topics, I've left out a lot from my discussion, both in the actual computations I explained and in possible extensions, related computations, and comparisons with real-world phenomena.  In particular:
 \begin{itemize}
  \item What happens when you include back-reaction of the gauge field on the geometry?
  \item The sudden approximation is not controlled by a small parameter (except maybe in some corners of $k$-space).  How close is it to the true $G^\Psi(k)$?
  \item How does $G^\Psi(k)$ change as we go from $T=0$ to $T=T_c$ for superconductivity?
  \item Little seems to depend on the choice of gauge group.  How about using $SO(4)$ with a fermion in the vector ${\bf 4}$ representation?  Is there any relation to the $SO(4)$ symmetry of the Hubbard model on a bipartite lattice?
  \item The branch cut structure is already visible in the strict IR limit.
  \item Poles in $G^\Psi(k)$ correspond to normal modes where $\psi \to 0$ both for $z \to 0$ and $z \to \infty$.
  \item Normal modes are restricted to the ``preferred region'' where $K_m^{\rm IR}$ is spacelike but $K_m^{\rm UV}$ is timelike.
  \item There's a recent extension to a $d$-wave condensate.\cite{Benini:2010qc}  This work also has the phenomenological advantage that it produces highly anisotropic Dirac cones.
  \item I've omitted discussion of the significant literature on
   \begin{itemize}
    \item Fermions in $s$-wave holographic superconductors.
    \item Conductivity at finite frequency.
    \item Thermodynamic and hydrodynamic properties of $p$-wave superconductors.
    \item Embedding holographic superconductors in string/M-theory.
   \end{itemize}
  \item Comparison of $G^\Psi(k)$ to results of ARPES measurements is interesting: You get a peak-dip-hump structure from combination of the normal mode and the continuum from inside the Dirac cone.
  \item Why not do a fermion response calculation for spin-$3/2$ fermions, for example the gravitini in actual supergravity theories?
  \item There are instabilities of holographic superconductors besides the ones that spontaneously break $U(1)$, for example the Gregory-Laflamme instability and runaways in moduli space.  How do all these instabilities compete?
 \end{itemize}
In short, it seems that there is still a lot to learn about holographic superconductors.

\section*{Acknowledgments}

I thank my collaborators: Silviu Pufu, Fabio Rocha, and Amos Yarom.  This work was supported in part by the Department of Energy under Grant No.~DE-FG02-91ER40671.

\bibliographystyle{ws-rv-van}
\bibliography{tasi10}

%\printindex[aindx]                 % to print author index
\printindex                         % to print subject index
\end{document}